\documentclass[twoside,11pt]{article}

\usepackage{melba}

\usepackage{amsmath,amsfonts}

\usepackage[nolist]{acronym}
\usepackage{caption}
\usepackage{subcaption}
\usepackage{graphicx}
\usepackage{bm}
\usepackage{cleveref}
\usepackage{float}
\usepackage{authblk}

\newcommand{\eac}[1]{\emph{\ac{#1}}}
\newcommand{\eacp}[1]{\emph{\acp{#1}}}
\newcommand{\eacf}[1]{\emph{\acf{#1}}}

\ShortHeadings{Framing image registration as a landmark detection problem (HitR)}{Waldmannstetter et al.}

\title{Framing image registration as a landmark detection problem for label-noise-aware task representation (HitR)}

\author[1,2,14]{\textbf{Diana Waldmannstetter}}
\author[1,5]{\textbf{Ivan Ezhov}}
\author[3]{\textbf{Benedikt Wiestler}}
\author[4]{\\\textbf{Francesco Campi}}
\author[4]{\textbf{Ivan Kukuljan}}
\author[6]{\textbf{Stefan Ehrlich}}
\author[7]{\\\textbf{Shankeeth Vinayahalingam}}
\author[8]{\textbf{Bhakti Baheti}}
\author[9]{\textbf{Satrajit Chakrabarty}}
\author[8]{\\\textbf{Ujjwal Baid}}
\author[8]{\textbf{Spyridon Bakas}}
\author[3]{\textbf{Julian Schwarting}}
\author[3]{\textbf{Marie Metz}}
\author[3]{\\\textbf{Jan S. Kirschke}}
\author[10,11]{\textbf{Daniel Rueckert}}
\author[12]{\textbf{Rolf A. Heckemann}}
\author[4]{\\\textbf{Marie Piraud}}
\author[2,13]{\textbf{Bjoern H. Menze}}
\author[1,3,4,5,13]{\textbf{Florian Kofler}} 

\affil[1]{\emph{Department of Computer Science, TUM School of Computation, Information and Technology, Technical University of Munich, Munich, Germany}}
\affil[2]{\emph{Department of Quantitative Biomedicine, University of Zurich, Zurich, Switzerland}}
\affil[3]{\emph{Department of Diagnostic and Interventional Neuroradiology, School of Medicine, Klinikum rechts der Isar, Technical University of Munich, Germany}}
\affil[4]{\emph{Helmholtz AI, Helmholtz Munich, Neuherberg, Germany}}
\affil[5]{\emph{TranslaTUM - Central Institute for Translational Cancer Research, Technical University of Munich, Germany}}
\affil[6]{\emph{Healthcare Technologies, SETLabs Research, Munich, Germany}}
\affil[7]{\emph{Department of Oral and Maxillofacial Surgery, Radboud University Nijmegen Medical Centre, Nijmegen, the Netherlands}}
\affil[8]{\emph{Division of Computational Pathology, Department of Pathology and Laboratory Medicine, School of Medicine, Indiana University, Indianapolis, IN, USA}}
\affil[9]{\emph{Department of Electrical and Systems Engineering, Washington University in St.Louis, St. Louis, MO, USA}}
\affil[10]{\emph{Artificial Intelligence in Healthcare and Medicine, Technical University of Munich, Munich, Germany}}
\affil[11]{\emph{Department of Computing, Imperial College London, London, U.K.}}
\affil[12]{\emph{Department of Medical Radiation Sciences, University of Gothenburg, Gothenburg, Sweden}}
\affil[13]{\emph{contributed equally as senior authors}}
\affil[14]{\emph{diana.waldmannstetter@tum.de}}

\begin{document}

\begin{acronym}
\acro{BraTS-Reg}{Brain Tumor Sequence Registration Challenge}
\acro{CNN}{Convolutional Neural Network}
\acro{TRE}{Target Registration Error}
\acro{IRR}{Interrater Reliability}
\acro{ROI}{Region of Interest}
\acro{HitR}{Landmark Hit Rate}
\end{acronym}

\maketitle

\newpage
\begin{abstract}
Accurate image registration is pivotal in bio-medical image analysis, where selecting suitable registration algorithms demands careful consideration.
While numerous algorithms are available, the evaluation metrics to assess their performance have remained relatively static.
This study addresses this challenge by introducing a novel evaluation metric termed \eac{HitR}, which focuses on the clinical relevance of image registration accuracy.
Unlike traditional metrics such as Target Registration Error, which emphasize sub-resolution differences, \eac{HitR} considers whether registration algorithms successfully position landmarks within defined confidence zones.
This paradigm shift acknowledges the inherent annotation noise in medical images, allowing for more meaningful assessments.
To equip \eac{HitR} with label-noise-awareness, we propose defining these confidence zones based on an Inter-rater Variance analysis.
Consequently, hit rate curves are computed for varying landmark zone sizes, enabling performance measurement for a task-specific level of accuracy.
Our approach offers a more realistic and meaningful assessment of image registration algorithms, reflecting their suitability for clinical and biomedical applications.
\end{abstract}

    


\section{Introduction}
Image registration is a fundamental task in medical image analysis and the amount of available registration methods is growing steadily.
Choosing an appropriate algorithm with a good balance between sufficient registration accuracy and reasonable computation time requires detailed consideration.
It is therefore important to find reliable validation metrics for being able to evaluate registration methods properly and assess their suitability for a specific task.
However, while developing new registration algorithms is ongoing, evaluation metrics have not changed much over the years. 
Recent works use either segmentation-based measures such as Dice \citep{balakrishnan2019voxelmorph} or landmark-based metrics calculating a \eac{TRE} \citep{baheti2021brain,hering2022learn2reg}.

\subsubsection*{Motivation and Contribution}
Well-performing algorithms are typically distinguished by small \eac{TRE} differences, often in a sub-resolution range.
For the interpretation of these \eacp{TRE}, it is crucial to keep inter- and intra-rater-reliability in mind to differentiate between actual performance improvements and random improved fits of the test set \citep{kofler2022approaching}.
\eac{HitR} draws its inspiration from valuable insights obtained through qualitative interviews with clinical experts.
These experts prioritize precise landmark localization, with a greater focus on avoiding major omissions rather than minor positional fluctuations.
They acknowledge that annotations inherently possess some level of inaccuracy, which shifts their focus on the identification of missed landmarks.

Consequently, we leverage these insights to redefine the evaluation of image registration algorithms through the introduction of \eac{HitR} metric.
Unlike traditional metrics that emphasize sub-resolution differences, \eac{HitR} is designed to align with clinical priorities by evaluating the ability of registration algorithms to position landmarks within predefined confidence zones.
We further enhance \eac{HitR}'s practicality by proposing a method for defining these zones based on Inter-rater Variance analysis.
Additionally, we compute hit rate curves for various zone sizes, allowing task-specific accuracy assessments.
This shift in paradigm offers a pragmatic approach for assessing image registration algorithms, aligning them more closely with clinical and biomedical applications.
We hope our work can help bridging the gap between algorithm development and real-world requirements.

\section{Related Work}
Image registration is an essential step for segmentation algorithms, either as part of the preprocessing pipeline \citep{kofler2020brats,buchner2023development,kofler2023we} or inside the algorithm itself \citep{fidon2022dempster}.
Moreover, image registration is used for pre-alignment of images in the scope of tumor growth models \citep{lipkova2019personalized,ezhov2023learn}.

In general, evaluation methods for measuring the quality and accuracy of image registration methods can be divided into two categories, qualitative and quantitative validation \citep{pluim2016truth}.
Qualitative evaluation usually involves assessment performed by (clinically) trained experts such as radiologists, judging the registration results by visual perception.
In quantitative registration evaluation, there is the option to define a reference transformation, which serves as a reference when evaluating registration accuracy.
Nevertheless, defining realistic and therefore "correct" reference deformations is challenging.
Hence, those kinds of validation can hardly be used for large datasets as they are predominantly used in image registration studies and challenges.
Quantitative validation metrics in the form of aligning segmentations and/or landmarks are usually chosen when it comes to evaluating registration algorithms to a greater extent \citep{hellier2003retrospective,murphy2010evaluation,marstal2019continuous,hering2022learn2reg,weitz2022acrobat,borovec2020anhir}.
There are predominantly two main methods for quantitative evaluation that are used in image registration studies and challenges \citep{pluim2016truth}: (i) alignment of corresponding anatomical structures (segmentations) and (ii) alignment of corresponding point sets (landmarks).
Both contour/surface alignment via segmentation and landmark alignment using annotated points in an image require human (expert) knowledge and are frequently performed manually.

Recent image registration studies and challenges along with the chosen evaluation metrics are described subsequently:

\begin{itemize}
    \item \emph{Retrospective Evaluation of Inter-subject Brain Registration} \citep{hellier2003retrospective} -- Registration of brain MR images: corresponding point set evaluation using \eac{TRE} and Hausdorff distance (HD); segmentation evaluation using Dice Similarity Coefficient (DSC), Union Coefficient (UC), Jaccard Coefficient (JC), False Negative Error (FNE), False Positive Error (FPE), and Volume Similarity (VS).
    \item \emph{The EMPIRE10 Study} \citep{murphy2010evaluation} -- Registration of intra-patient thoracic CT images: alignment of lung boundaries; alignment of major fissures; evaluation of corresponding landmarks (\eac{TRE}); physical plausibility of registration deformation (folding/tearing).
    \item \emph{The Continuous Registration Challenge} \citep{marstal2019continuous} -- Registration of lung CT and brain MR images: tissue classification by texture analysis; extraction of differential characteristics; distance between extracted cortical sulci.
    \item \emph{Learn2Reg Challenge} \citep{hering2022learn2reg} -- Registration of inter- and intra-patient images from different modalities (ultrasound, CT, MR) for multiple anatomies (brain, abdomen, thorax: segmentation-based alignment evaluation (DSC, HD); corresponding landmark evaluation (\eac{TRE}); plausibility (smoothness) of displacement field (Jacobian determinant).
    \item \emph{ANHIR Challenge} \citep{borovec2020anhir} -- Registration of histology (microscopy) images: corresponding landmark evaluation (\eac{TRE}).
    \item \emph{ACROBAT Challenge} \citep{weitz2022acrobat} -- Registration of histopathological whole slide images (WSI) with immunohistochemically (IHC) stained breast cancer tissue sections: corresponding landmark evaluation (\eac{TRE}).
\end{itemize}

Nevertheless, \eac{TRE} has several limitations \citep{luo2022dataset}: Landmarks are annotated by localization algorithms (manual, automatic or semi-automatic methods) and may contain localization errors.
Moreover, \eac{TRE} only estimates the error at specific landmark locations.
Hence, a dense population of landmarks is preferable since it is enabling registration accuracy evaluation on the whole image. 
Also, sparse or unequally distributed landmarks may introduce bias.
At the same time, dense point sets are costly, especially when the annotations are performed manually by human experts.

Therefore, we aim to move towards a more label-noise-aware representation instead of solely relying on single \eac{TRE} for evaluating image registration algorithms.

\section{Methods}
\label{sec:methods}
Building on the foundation of the \eac{TRE}, we address the challenge of managing multiple annotators and their potential biases by computing the average annotated landmark.
This leads to the introduction of \eac{HitR}, a landmark-based hit-rate metric, which evaluates prediction accuracy within predefined \eacp{ROI} around landmarks.
For better visualization, we propose a method for extrapolating radii that is particularly useful for landmarks with limited annotations.
Lastly, we explore the option of aggregating distances at the image domain level, addressing variations between images and multiple annotations.

\subsection{Target Registration Error and Euclidean distance}
Registration algorithms are frequently evaluated based on landmarks by computing the \eacf{TRE}.
The \eac{TRE} is typically expressed as the Euclidean distance between the reference landmark $ \bm{y} $ and the model prediction $ \hat{\bm{y}} $:

\begin{equation}
    TRE(\bm{y} , \hat{\bm{y}}) =
    \lVert \bm{y} - \hat{\bm{y}} \rVert_2 .
    \label{eq:TRE}
\end{equation}

Specifically, the Euclidean distance between two vectors $ \bm{p}, \bm{q} \in \mathbb{R}^d $ is defined as:

\begin{equation}
    \lVert \bm{p} - \bm{q} \rVert_2  =  \sqrt{\sum_{i=1}^{d} (p_i - q_i)^2},
    \label{eq:euclidean}
\end{equation}

where $p_i $ and $ q_i $ denote the $i$-th components of the vectors (i.e. $ \bm{p} = (p_1, p_2, \dots, p_n) $).

\subsection{Accounting for multiple annotation entities}
To account for annotation noise induced by the annotation entities (from now on called \emph{annotators}) and their respective biases, we can calculate the average annotated landmark in space:

\begin{equation}
    \bm{y}^{(k,l)} = \frac{1}{n_{k,l}} \sum_{v = 1}^{n_{k,l}} \bm{y}^{(k,l,v)},
    \label{eq:ykl}
\end{equation}

where $\bm{y}^{(k,l,v)} $ is the annotation for the landmark $l$ in the image domain $k$ for the annotator $v$, and $\bm{y}^{(k,l)} $ is the average annotation that we use as a reference to compare with model predictions.
Additionally, the number of annotators $n_{k,l}$ depends on the image domain $k$ and the landmark $l$ to address the possibility of having different annotators for different landmarks.

Optionally, \Cref{eq:ykl} can be extended to account for varying levels of annotator reliability by computing a weighted average of the landmark annotations:
\begin{equation}
    \bm{y}^{(k,l)} =\sum_{v = 1}^{n_{k,l}} \lambda_v \bm{y}^{(k,l,v)}, \ s.t. \ \sum_{v} \lambda_v = 1,
    \label{eq:ykl_with_lambda}
\end{equation}

where the weights $ \lambda_v $ represent the reliability of the annotator $ v $.

Now we can calculate each annotator's distance to the average landmark and define a set $ D_{k,l} $ containing these distances:
\begin{equation}
    D_{k,l} = \left\{ \lVert \bm{y}^{(k,l,v)} - \bm{y}^{(k,l)} \rVert_2 \middle| \, v = 1, 2, \ldots, n_{k,l}\right\},
    \label{eq:Dkl}
\end{equation}

where $ l $ represents the landmark and $ k $ the image and $ v $ the different annotators.

Next, we combine all these sets across image domains and landmarks to obtain a set $ D $ containing all the annotation distances:
\begin{equation}
    D = \bigcup_{k = 1}^{K} \bigcup_{l = 1}^{L_k} D_{k,l},
    \label{eq:D}
\end{equation}

where $ K $ is the total number of images and $ L_k $ the number of landmarks in image $ k $.
Similar to \Cref{eq:ykl}, $ L_k $ depends on the image domain $k$ as the total amount of landmarks can vary across images.

\subsection{Computing the landmark-based hit-rate (accuracy)}
\label{sec:hitrate}
For most registration problems, tiny variations around the reference landmarks are tolerable and expected, for instance, they naturally occur due to transformation artifacts when re-sampling from different resolutions.
Consequently, sub-resolution \eacp{TRE} frequently possess questionable meaning and should be disregarded by practitioners.

Instead, practitioners typically care about the presence of substantial registration errors, in other words, missed landmarks.

Therefore, we construct a ball of radius $ r $ centered at each landmark, which we call \eac{ROI}, and we consider a prediction a \emph{hit} when it is located within the \eac{ROI} and a \emph{miss} otherwise.
Function $ \chi_r (\cdot, \cdot) $, as described in \Cref{eq:hits}, then assigns $ 1 $ for each hit and $ 0 $ for each missed landmark:

\begin{equation}
    \chi_r (\bm{y}, \hat{\bm{y}}) =
    \begin{cases}
        1, & \lVert\bm{y} - \hat{\bm{y}}\rVert_2 \leq r \\
        0, & \lVert\bm{y} - \hat{\bm{y}}\rVert_2 > r    \\
    \end{cases}.
    \label{eq:hits}
\end{equation}

\begin{figure}[H]
    \centering
    \includegraphics[width=\textwidth]{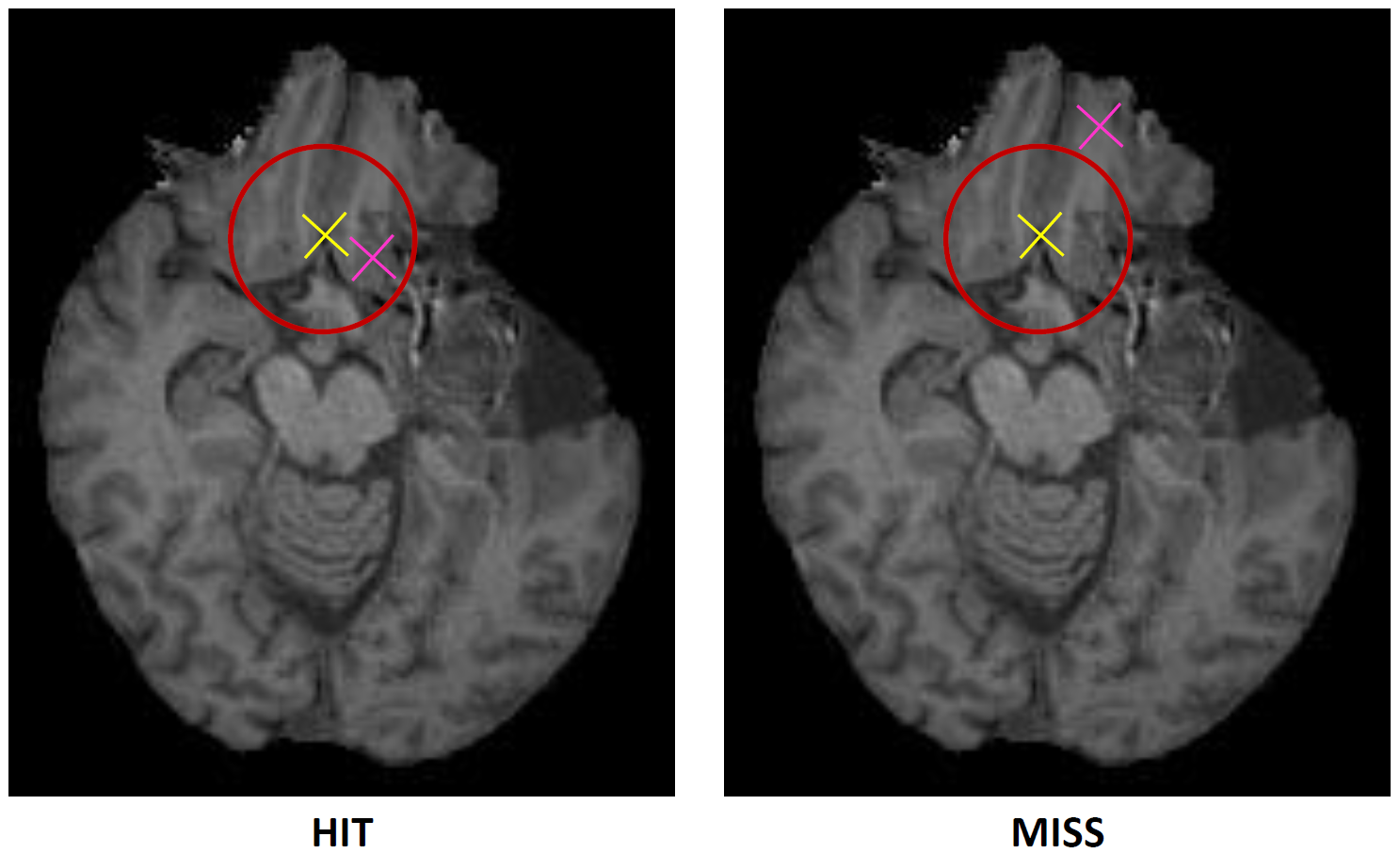}
    \caption{
        \emph{Hit} or \emph{Miss}.
        Reference annotation (yellow) and predicted landmark (magenta).
        The red circle with radius $ r $ indicates the tolerated \eac{ROI} around a landmark, where the radius corresponds to a specified threshold derived by annotators' distances.
        A hit is classified according to \Cref{eq:hits} and the hit rate (\eac{HitR}) is calculated with \Cref{eq:metric}, describing the ratio between hits and number of landmarks.
    }
    \label{fig:hit_miss}
\end{figure}

\Cref{fig:hit_miss} showcases a 2D illustration of this process.
As input for radius $ r $, we supply the distances derived from \Cref{eq:Dkl}.

Let's consider an image domain $k$ with landmarks $l = 1, \dots, L_k$.
We define the sets containing the positions of the reference landmarks as $ Y_k = \{ \bm{y}^{(k,l)}\}_{l=1}^{L_k}$ and the predicted landmarks as $ \hat{Y}_k = \{\hat{\bm{y}}^{(k,l)}\}_{l=1}^{L_k}$, where the components $\bm{y}^{(k,l)}$ and $\hat{\bm{y}}^{(k,l)}$ represent the reference prediction and the model prediction for a landmark $l$ respectively.
Now the landmark-based hit rate, or in other words, the \emph{accuracy} in the image domain $k$ in dependency of radius $ r $ is computed as the ratio between the hits and the total number of landmarks:
\begin{equation}
    m_r(Y_k, \hat{Y}_k) = \frac{1}{L_k}\sum_{l = 1}^{L_k} \chi_r (\bm{y}^{(k,l)}, \hat{\bm{y}}^{(k,l)}).
    \label{eq:metric}
\end{equation}

\newpage
\subsection{Extrapolation of radii and visualization}
\label{sec:visualization}
To achieve a \emph{gold} standard, the whole dataset should be annotated by several annotators.
As this will often turn out to be prohibitively expensive, we propose to approximate this \emph{gold}- with a \emph{silver} standard.
For this, it is necessary to have a representative subset of the dataset re-annotated by at least one annotator and compute $ D $ (cf. \Cref{eq:D}).

Now, instead of directly plugging the distances from $ D $ into the metric calculation (cf. \Cref{eq:metric}), we can sample from the distribution of $ D $ and extrapolate the radii for landmarks with only one annotation.
For sampling radii, we propose the following approach:
\begin{equation}
    r_{\mu} = \text{median}(D) + \mu \text{MAD}(D).
    \label{eq:sampling}
\end{equation}

where MAD is the median absolute deviation.

We propose this approach as the median and median absolute deviation are robust towards outliers and promise to generate meaningful intervals for most distributions.

Additionally, we can fine-tune $ \mu \in [\mu_{min}, \mu_{max} ]$  to cover a desired range of accuracy while maintaining a non-negative radius ($ r \geq 0 $).

For visualization, we propose to plot accuracy curves dependent on the radius $ r $ as illustrated in \Cref{fig:linechart_hitr}.



\subsection{Aggregation on the level of the image domain}
For some use cases, it can make sense to aggregate the distances on the level of the image domain.
For instance, if there is a high variance between images and multiple annotations exist for a subset of the landmarks within an image $ k $.
In this case \Cref{eq:Dkl} can be substituted with:
\begin{equation}
    D_k = \bigcup_{l = 1}^{L_k} D_{k,l}
    \label{eq:Dk}
\end{equation}

which leads to substituting the approximation in \Cref{eq:sampling} with:
\begin{equation}
    r_{\mu, k} = \text{median}(D_k) + \mu \text{MAD}(D_k)
    \label{eq:r_mu_k}
\end{equation}

where MAD is the median absolute deviation.

Then the accuracy (cf. \Cref{sec:hitrate}) can be computed analogously.

\newpage
\section{Experiments}
To evaluate and showcase the capabilities of \eac{HitR} we conduct experiments on the dataset of the \eac{BraTS-Reg} \citep{baheti2021brain}.

\subsection{BraTS-Reg Challenge}
\label{sec:bratsreg}
\eac{BraTS-Reg} focuses on aligning post-tumor-resection MR images with their pre-surgery counterparts in brain MR scans and can be characterized as follows.

\subsubsection*{Dataset}
The multi-institutional dataset used for evaluation consists of 50 paired quartets of baseline (pre-operative) and follow-up (post-operative) brain MR images.
For each time point, the four MR sequences native T1-weighted (T1), contrast-enhanced T1 (T1-CE), T2-weighted (T2), and T2 Fluid Attenuated Inversion Recovery (FLAIR) are provided, respectively.
All images are preregistered to the same anatomical space.

\subsubsection*{Annotations}
Several experts from multiple institutions annotated 739 pairs of corresponding anatomical landmarks in the baseline and the follow-up images.
The number of annotations varies for each case, ranging from 6 to 50 landmarks for one patient.
For each image, all provided annotations are from the same expert.
Annotation variance was assessed on post-surgery landmarks for a subset of the patients.

\subsubsection*{Challenge Submissions}
15 different algorithms, covering a wide range of registration methods, including iterative and Deep Learning-based approaches, were submitted to \eac{BraTS-Reg}.

\subsubsection*{Evaluation Metric}
\eac{BraTS-Reg} was evaluated mainly by calculating the \eac{TRE} between landmarks warped by the participating registration algorithms and the respective reference annotations.

\newpage
\subsection{Simulation of Annotation noise}
\begin{figure}[H]
    \centering
    \includegraphics[width=1.0\textwidth]{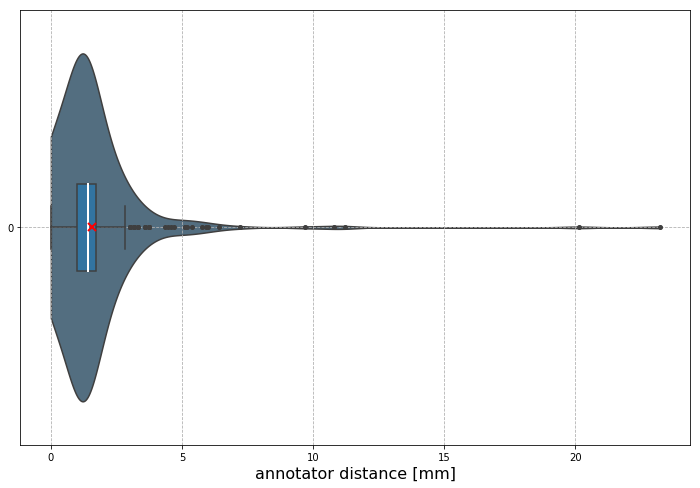}
    \caption{
        Distribution of distances between annotators' on a set of 399 re-annotated landmarks in the \eacf{BraTS-Reg}, see \Cref{sec:bratsreg}.
        Most landmarks are re-annotated landmarks within a distance of around $5mm$, while a few landmarks are re-annotated with distances up to almost $24mm$.
    }
    \label{fig:annotation_error}
\end{figure}

Even though \eac{BraTS-Reg} provided only landmarks from a single rater for most of the test set, it evaluates the inter-rater variance on a representative subset of post-surgery landmark annotations.
The distribution of the differences between annotators is illustrated in \Cref{fig:annotation_error}.

We utilize this distribution to simulate the behaviors of 20 virtual annotators.
For this, we introduce random perturbations to the 739 landmarks of the reference annotation within 3D space.
For each dimension $ (x, y, z) $, we stochastically sample a value from the distribution representing annotation differences.
Consequently, we multiply the sampled value by an annotator-specific bias $ \beta $, randomly selected from a uniform distribution spanning from $ 0.7 $ to $ 1.3 $.
This way we obtain new landmark annotations with similar characteristics to the original labels.
Subsequently, we apply the resulting product to the respective x, y, and z dimensions of the landmark, thus creating a new simulated landmark in 3D space.
Code for simulating new landmarks is provided \emph{\href{https://pastebin.com/aEpJaZAQ}{here}}.

\clearpage
\subsection{HitR Analysis}
\begin{figure}[H]
    \centering
    \includegraphics[width=\textwidth]{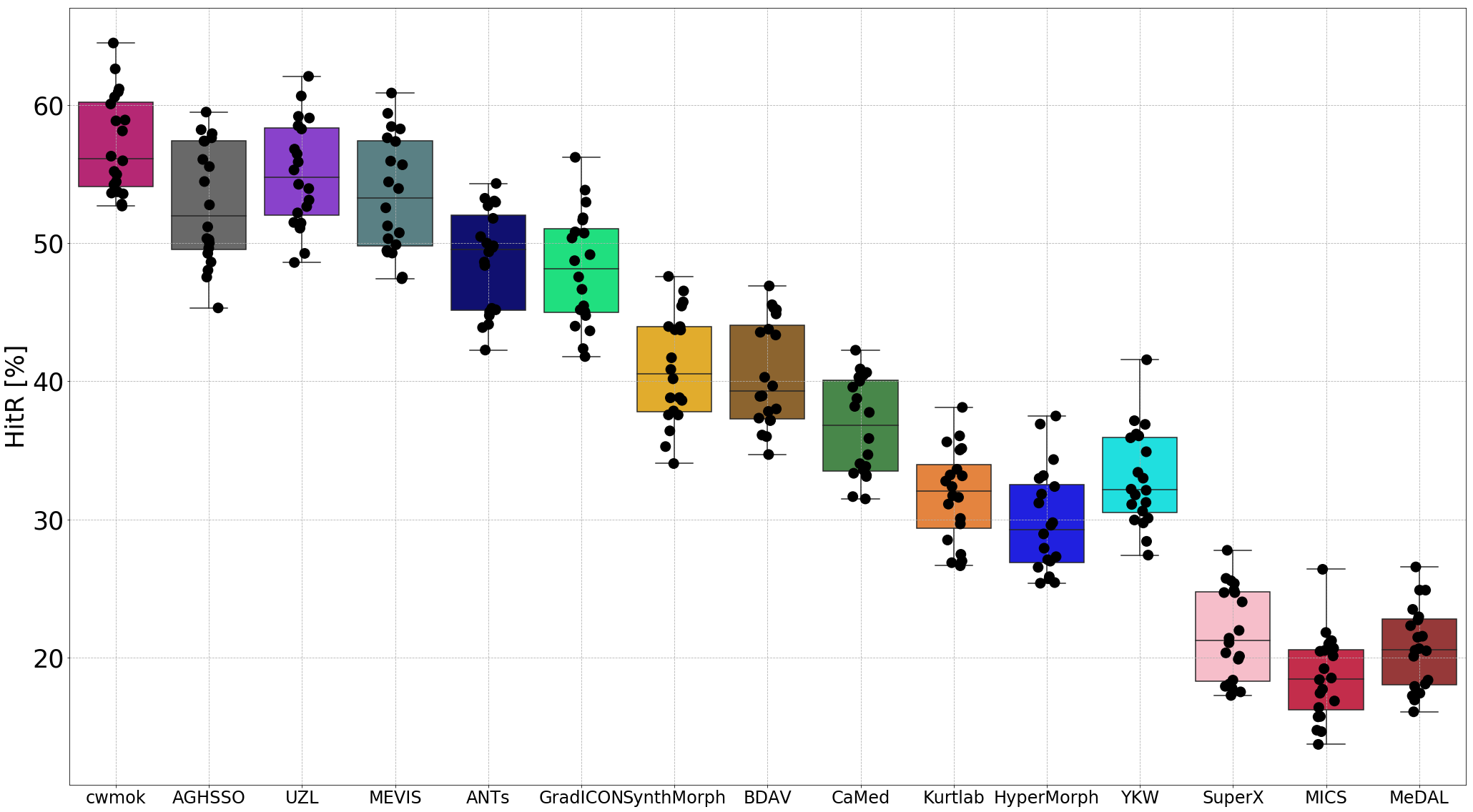}
    \caption{
        Boxplot illustrating \eac{HitR} per registration and annotator.
        \eac{HitR} is computed according to \Cref{eq:metric}.
        For the computation, we consider each rater's annotations' distance to the average landmark as a radius for the \eac{ROI}.
        For each algorithm, \eac{HitR} is calculated based on all registered landmarks and the defined \eac{ROI}.
        The evaluated algorithms are submissions to the \eac{BraTS-Reg}, see \Cref{sec:bratsreg}.
        The performance of the algorithms significantly varies and overlaps depending on the underlying annotation confirming the decision of the challenge organizers to award the challenge contributions based on performance tiers.
    }
    \label{fig:hitrate_boxplot}
\end{figure}

The simulated annotations enable us to apply the methodology developed in \Cref{sec:methods} and compute the \eac{HitR}.
Consequently, we start by computing the average annotated landmark as described in \Cref{eq:ykl}.
This enables us to compute the \eac{HitR} per registration algorithm and annotator, following \Cref{eq:metric}, as illustrated in \Cref{fig:hitrate_boxplot}. 
We can observe that algorithms' performances differ depending on the respective landmark annotation by varying raters.

\begin{figure}[H]
    \centering
    \includegraphics[width=\textwidth]{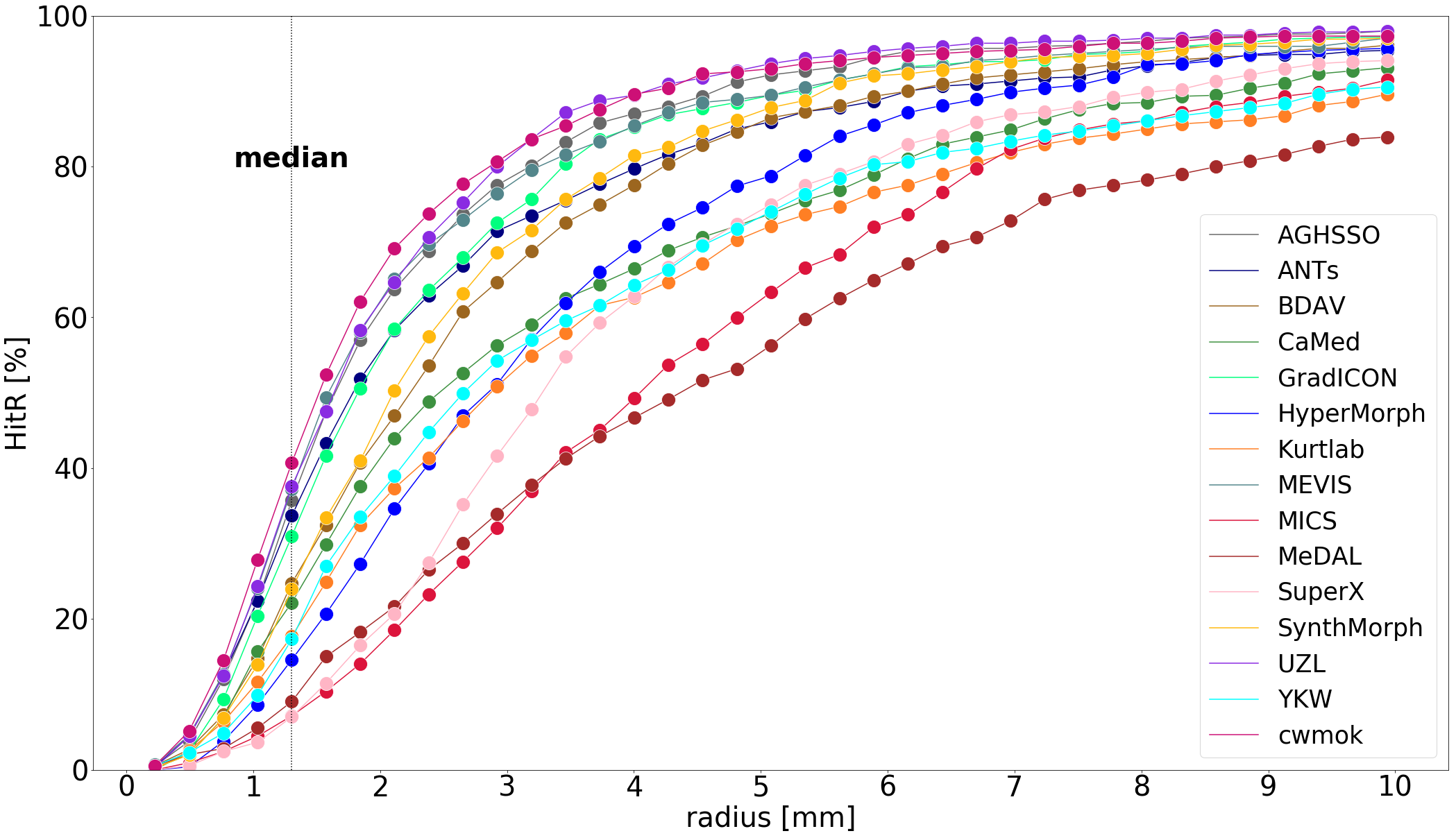}
    \caption{
        Line chart comparing \eac{HitR} in dependency of the \emph{radius} $r$ for various registration algorithms.
        The evaluated algorithms are submissions to the \eac{BraTS-Reg}, see \Cref{sec:bratsreg}.
        \eac{HitR} is computed for the points derived from \Cref{eq:sampling}, while the lines are interpolated.
        Some algorithms reveal greatly improved performance once the \eac{ROI} size is increased (indicated by crossing lines).
    }
    \label{fig:linechart_hitr}
\end{figure}

Further, we can compute \eac{HitR} assuming the same \emph{radius} $ r $ across all landmarks, simulating a scenario where we don't have annotations for the whole dataset.
For this, we compute varying radii as described in \Cref{sec:visualization}.
\Cref{fig:linechart_hitr} illustrates the behavior of algorithms when radii are sampled from the distribution of distances between annotators (see \Cref{fig:annotation_error}) using \Cref{eq:sampling}. 
While top ranked algorithms behave rather stable over all applied radii, some of the evaluated algorithms perform vastly better when radii are increasing, implied by crossing curves. 
This finding can give hints towards the overall robustness of algorithms.

\subsection{Comparison to \eac{TRE}}
\begin{figure}[h]
    \centering
    \includegraphics[width=\textwidth]{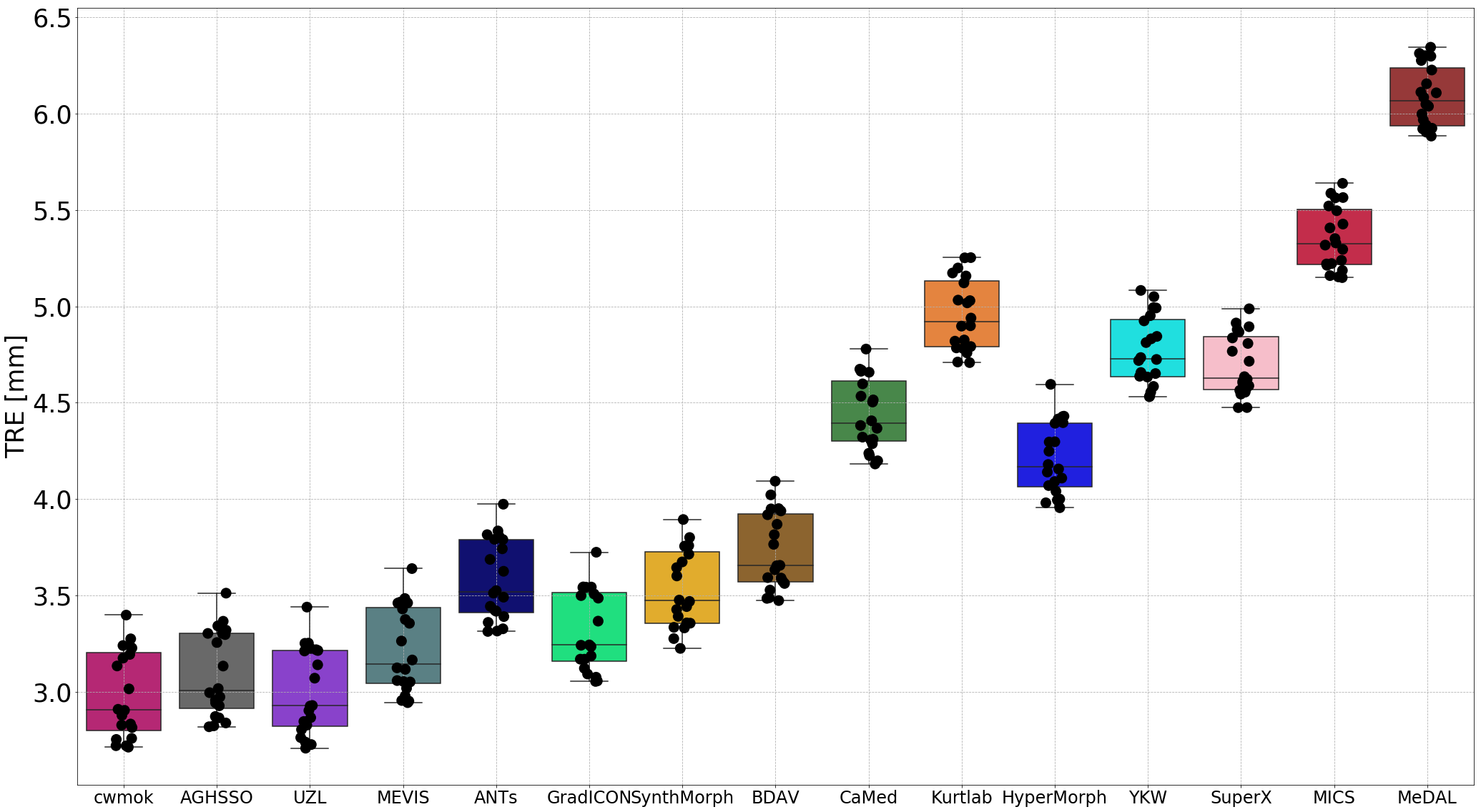}
    \caption{
        Boxplot illustrating \eac{TRE} to all annotators' landmark annotations per registration algorithm. 
        For each algorithm, the \eac{TRE} between a registered landmark and every corresponding annotators' landmark is computed, respectively.
        The evaluated algorithms are submissions to the \eac{BraTS-Reg}, see \Cref{sec:bratsreg}.
        With a Pearson $ r $ of $ -0.49 $, \eac{TRE} only moderately correlates with \eac{HitR}.
    }
    \label{fig:boxplot_tre}
\end{figure}

Next, we analyze how \eac{HitR} performs in comparison to \eacf{TRE}.
Therefore, we compare the hits and misses to the \eac{TRE} per landmark.
We observe a moderate Pearson $ r $ of $ -0.49 $.
This demonstrates that \eac{HitR} even though it is computed in dependency of \eac{TRE} measures something different.
\Cref{fig:boxplot_tre} illustrates \eac{TRE} across the test set.

\section{Discussion and Conclusion}
Our study introduces \eac{HitR}, a label-noise-aware metric designed to evaluate registration algorithms, with an illustrative application within the \eacf{BraTS-Reg}.
The metric holds potential for broader applicability in various registration problems within and beyond the bio-medical imaging realm.
Furthermore, our simulation study provides a robustness estimation of the \eac{BraTS-Reg} results by emulating real-world variations introduced by annotator biases and annotation errors.

\eac{HitR} allows explicit formulation of use-case-specific accuracy requirements (see \Cref{fig:linechart_hitr}) by defining fixed thresholds for \eac{ROI} size.
Depending on the downstream biomedical task, this can help decide between computational run-time and registration accuracy.
For example, in pre-registering multiple modalities (PET and MRI) for stereotactic biopsy planning, highly accurate registration is mandatory, but the run-time can be neglected.
On the other hand, when doing longitudinal registration for evaluation of tumor response, run-time will be prioritized (as this is done while reading the images as they are being acquired).
Here, the RANO criteria, which define evaluation guidelines for assessing brain tumor growth in MRI evaluation, allow for some relaxation in registration accuracy requirements due to the significant margin of growth required to diagnose tumor progression.
We hope that allowing such explicit formulation of use case requirements \eac{HitR} can contribute to advancing registration techniques towards improved applicability and accuracy.

\subsection*{Limitations and Future Research}
One limitation of our approach is that landmark \eacp{ROI} are modeled with ball shapes.
This is problematic as it assumes a homogeneous distribution of annotation noise.
In reality, annotation noise might be heterogeneously distributed, as in our practical example annotators might generate more noisy labels in tumor vicinity.
Future research could consider this by modeling this with more complex topologies.

Furthermore, the proposed metric is not differentiable and can hence not be applied to train \eacp{CNN}.
Instead of discretizing, future work could explore extending the metric to a differentiable form.
This could be achieved by measuring the signed distance from the landmark \eac{ROI} and thus enabling usage as a loss in \eac{CNN} training.

\clearpage

\acks{Supported by Deutsche Forschungsgemeinschaft (DFG) through TUM International Graduate School of Science and Engineering (IGSSE), GSC 81.
BM, BW and FK are supported through the SFB 824, subproject B12.
BM acknowledges support by the Helmut Horten Foundation.}

%


\bibliography{mybib}

\end{document}